\documentclass[prd,superscriptaddress,amsmath,notitlepage,twocolumn]{revtex4-2}
\usepackage{graphicx}
\usepackage{float}
\usepackage{color}
\usepackage{epstopdf,cancel}
\usepackage{epsf,latexsym,bbm,euscript}
\usepackage{amssymb,amsmath}
\usepackage{times,graphics}
\usepackage{soul,xcolor}
\usepackage{mathtools}
\usepackage{braket}
\usepackage{svg}
\usepackage{url,hyperref}
\hypersetup{colorlinks,linkcolor={blue!55!black},citecolor={red!45!black},urlcolor={blue!45!black},breaklinks=true}

\newcommand{\be}{\begin{equation}}
\newcommand{\ee}{\end{equation}}

\begin{document}

\title{The Hawking temperature of dynamical black holes via Rindler transformations}

\author{Pravin Kumar Dahal}
\email{pravin-kumar.dahal@hdr.mq.edu.au}

\author{Fil Simovic}
\email{fil.simovic@mq.edu.au}
\affiliation{School of Mathematical and Physical Sciences, Macquarie University, Sydney, NSW 2109, Australia}

\begin{abstract}

The Vaidya metric serves as a useful model-building tool that captures many essential features of dynamical and/or evaporating black hole spacetimes. Working in a semiclassical setting, we show that in the limit of slow evaporation, a general spherically symmetric metric subject to certain regularity conditions is uniquely described by a linear ingoing Vaidya metric in the near-horizon region. This suggests a universal description of the near-horizon geometry of evaporating black holes in terms of the linear Vaidya metric. We also demonstrate that the linear Vaidya metric can be brought into manifestly conformally static form, allowing us to determine the Hawking temperature associated with the Vaidya background with respect to the conformal vacuum. Since back-reaction is implicitly accounted for, we conclude that slowly evaporating black holes are indeed accurately described by quasistatic sequences of Schwarzschild metrics even when dynamical effects are present.

\end{abstract}

\maketitle

\section{Introduction}

Black holes, both as mathematical constructs and physical objects, have a long history of intrigue and speculation. Though their status as true astrophysical objects resulting from gravitational collapse remains speculative \cite{barack2019,cardoso2019,murk2023}, evidence for ultra-compact objects which are well described by classical black hole solutions of general relativity is abundant \cite{reynolds2003,bambi2017,ingram2019,eht2019,ligo2021}. While their physical reality remains somewhat uncertain, the mathematical notion of a black hole represents an invaluable theoretical tool, having precipitated numerous insights into the microscopic structure of gravity, quantum systems, and information theory \cite{carlip2014,maldacena2020}. Black holes have also been implicated in a variety astrophysical phenomena, including quasars, active galactic nuclei, and gamma-ray bursts, providing critical insights into the formation and evolution of galaxies \cite{heckman2014,villanueva-domingo2021}. As such, they continue to provide fertile ground for new discoveries and advancements in modern theoretical and observational physics alike.
\\

While static solutions of general relativity are the most well-understood, astrophysical black holes are almost certainly highly dynamical. Various models of both the formation and evaporation phases of black holes have been proposed, including those described by Vaidya metrics, regular black holes, thin shells, and others \cite{martin-dussaud2019,piesnack2022,bardeen1968a,carballo-rubio2018a,baccetti2019b,ho2020}. Many of these models have a domain of validity which covers only the quasistatic portion of the black hole evolution---it is assumed that evaporation happens slowly over most of the black hole's lifetime, and that the collapse process can be described well by semiclassical physics up to the Planck scale. Common both to derivations of Hawking radiation and the laws of black hole thermodynamics is that any back-reaction on the geometry can be ignored. While this seems plausible for black holes of observationally relevant sizes, a clear answer can only be provided by solving the semi-classical Einstein equations
\be
G_{\mu \nu}=\bra{\psi}\hat{T}_{\mu \nu}\ket{\psi}\ ,
\ee
for which even an iterative solution is currently out of reach.

One very useful proxy for representing the full dynamical spacetime of an evaporating black hole is the Vaidya metric \cite{vaidya1953} in ingoing null form
\be\label{met1}
ds^2=\left(1-\frac{2M(v)}{r}\right)dv^2+2dvdr+r^2d\Omega^2\ ,
\ee
where $v=t+r^*$ is defined through the Tortoise coordinate $r^*$ and $M(v)$ is the mass function. This metric describes the geometry associated with the spherically symmetric absorption of null dust (a pressureless radiation fluid) by a central mass. Heuristic pictures of Hawking radiation typically involve the presence of a negative energy flux near the horizon which serves to reduce the black hole's mass, and it is precisely this behaviour that is captured by \eqref{met1}. More sophisticated models of evaporating black holes can be constructed by patching together different metrics to appropriately capture various dynamical features which enter the description, as in the early model of Hiscock \cite{hiscock1981}, the Hayward model \cite{hayward2006b}, and others. While the Vaidya metric is both extremely useful and pervasive in the landscape of evaporation models, justification of its use is often tied to speculative intuition about the nature of Hawking radiation rather than arising as a demonstrable limit of some appropriate approximations. 
\\

The first element of our work addresses the validity of the Vaidya metric in describing the evaporation process. We will show that a minimal set of operationally motivated assumptions is sufficient to show that a generally spherically-symmetric dynamical metric limits to the ingoing Vaidya metric near the apparent horizon. This helps to justify its use in modelling important features of evaporating black holes. While elements of such a justification can be found in the existing literature, the explicit metric-limiting behaviour and the role of regularity is not discussed in full. The next issue which immediately arises in dynamical backgrounds like Vaidya is the lack of a globally timelike Killing vector field with respect to which one can decompose modes into their positive and negative frequency components, precluding the definition of a unique vacuum state. When symmetries are present, however, it is often possible to identify a ``natural" choice of vacuum state such that the ground state is invariant under the same symmetries as the underlying spacetime. For conformally flat spacetimes (those which are conformally related to Minkowski space)  there is such a preferred notion of vacuum state---the conformal vacuum. That there is a unique ground state which is invariant under conformal transformations allows one to compute renormalized expectation values of quantities like $\braket{T_{\mu\nu}}$ and $\braket{\phi^2}$ in conformally related spacetimes. 
\\

In this work, we introduce a particular conformal transformation that allows us to exploit the relationship between field theories defined on conformally related geometries to determine the temperature of Hawking radiation in the Vaidya background. Our result is important not only for enabling generalizations of the laws of black hole thermodynamics to dynamical backgrounds (for which there exist a number of proposals \cite{hayward1998b,booth2004,mishra2018}) but also in demonstrating that an evaporating black hole can indeed be adequately represented by a quasistatic sequence of Schwarzschild black holes with decreasing mass. That this is the case indicates that neglecting the back-reaction of the evaporation process on the geometry is justified in the semiclassical setting, despite existing suggestions that this might not be the case, and a lack of complete solution to the back-reaction problem.
\\

This paper is organized as follows: In Section \ref{sec2} we describe how regularity considerations lead to an ingoing Vaidya metric for the near-horizon region of generic slowly evaporating black hole spacetimes. In Section \ref{sec3}, we implement a series of coordinate transformations which bring the Vaidya metric into manifestly conformally static form. In Section \ref{sec4}, we exploit the conformal relationship between the Vaidya metric and the Rindler metric to compute the temperature associated with Hawking radiation in the Vaidya background. We also discuss how our results are consistent with  other methods used to extract thermal spectrums in black hole backgrounds. In Section \ref{sec5} we extend the result to Kerr-Vaidya spacetimes. WE summarize our results in Section \ref{sec6} and suggest directions for future work. Unless otherwise noted, we work in units where $\hbar=c=G=k_\text{B}=1$ and use Latin indices for spatial coordinates.

\section{Near-horizon geometry and Vaidya}\label{sec2}
We begin by demonstrating that the near-horizon geometry of a slowly evaporating black hole is given by the ingoing Vaidya metric, by starting with a general spherically-symmetric geometry in ingoing null coordinates. Such a metric can be written as
\be\begin{aligned}\label{met}
ds^2=-e^{2\phi(v,r)}&f(v,r)dv^2+2e^{\phi(v,r)}dv\,dr+r^2d\Omega^2\ ,\\
f(v,r)\equiv 1&-\dfrac{2M(v,r)}{r}\ ,
\end{aligned}\ee
where $\phi(v,r)$ is an integrating factor appearing in coordinate transformations and $d\Omega^2$ is the metric on $\mathcal{S}^2$. The apparent horizon is the (timelike) surface $g_{vv}=0$ which occurs at $r_{AH}=2M(v,r)$. From \eqref{met}, outgoing radial null geodesics obey
\be\label{drdv}
\dfrac{dr}{dv}=\frac{1}{2}e^{\phi(v,r)}\left(1-\dfrac{2M(v,r)}{r}\right)\ ,
\ee
where the $(v,r)$ dependencies will be suppressed henceforth. Using the fact that
\be
\dfrac{d}{dv}=\partial_v+\frac{1}{2}e^{\phi}\left(1-\frac{2M}{r}\right)\dfrac{\partial}{\partial r}=\dfrac{\partial}{\partial v}+\dfrac{dr}{dv}\dfrac{\partial}{\partial r}\ ,
\ee
one can compute
\be\label{eh}
\dfrac{d^2r}{dv^2}=\dfrac{\partial\phi}{\partial r}\left(\dfrac{dr}{dv}\right)^2+\left(\dfrac{\partial\phi}{\partial v}-\dfrac{e^\phi}{r}\dfrac{\partial M}{\partial r}+\dfrac{M\,e^\phi}{r^2}\right)\dfrac{dr}{dv}-\dfrac{e^\phi}{r}\dfrac{\partial M}{\partial v}
\ee
where no approximations have been made thus far. The Einstein equations for the metric \eqref{met} are
\begin{align}
4\pi r^2\, {T^v}_v&=-\dfrac{\partial M}{\partial r}\label{efe1} \\
4\pi r \, T_{rr}&=\dfrac{\partial \phi}{\partial r} \label{efe2}\\
4\pi r^2\, {T^r}_v&=\dfrac{\partial M}{\partial v} \label{efe3}
\end{align}
where the components of $T_{\mu\nu}$ are taken to be components of the expectation value of some suitably renormalized stress-energy tensor $\braket{T_{\mu\nu}}$ in the semiclassical approximation. Assume this stress-energy tensor near the horizon is responsible for the observed radiative flux at infinity. Conservation of energy $\nabla_{\mu}{T^{\mu}}_{\nu}=0$ combined with the Einstein equations allows one to relate the luminosity observed in the far region to the stress-energy tensor as $L\approx \partial_v M$, which is then consistent with the observed momentum flux ${T^r}_v=L/4\pi r^2$ at finite distance $r$ from the black hole (a prime indicates a derivative with respect to the argument). 
\\

We now assume that the luminosity is small and work to leading order in $L$ or $\partial_v M$. The event horizon is null to leading order in $L$ and at the (approximate location of the) event horizon outgoing null rays will be at rest not just momentarily (as on the apparent horizon) but for a timescale long compared to $M$. The condition is that
\be
\dfrac{d^2r}{dv^2}\bigg|_{r_h}=0\ , \label{sp11}
\ee
where $r_H$ is the location of the event horizon. Using \eqref{eh} one has, to leading order in $L= \partial_v M$ and near $r=2M$ that:
\begin{align*}
\dfrac{d^2r}{dv^2}&\approx\dfrac{1}{4}\left(1-\dfrac{2M}{r}\right)^2\dfrac{\partial\phi}{\partial r}+\dfrac{1}{2}\left(1-\dfrac{2M}{r}\right)\left(\dfrac{\partial\phi}{\partial v}+\dfrac{M}{r^2}\right)\nonumber\\
&\quad-\dfrac{1}{r}\dfrac{\partial M}{\partial v}\\
&\approx \dfrac{M(v)}{2r^2}\left(1-\dfrac{2M(v)}{r}\right)-\dfrac{M'(v)}{r}=0
\end{align*}
In the first line, \eqref{drdv} is used with $\phi= 0$ near $r=2M$, and $M(v,r)$ is taken to be approximately spatially uniform. In the second line, the sub-leading terms $(1-2M/r)^2 \partial_r \phi$ and $(1-2M/r) \partial_v \phi$ are discarded. These assumptions follows naturally from the regularity condition: curvature invariants constructed from components of $T_{\mu\nu}$ should not diverge on the apparent horizon, otherwise the horizon structure itself becomes tenuous (details are provided in the Appendix). The event horizon is thus approximately located at
\be\label{reh1}
r_H=\dfrac{M(v)-\sqrt{M(v)^2-16M(v)^2M'(v)}}{4M'(v)}\ ,
\ee
which to leading order in $L=M'(v)$ gives
\be\label{reh2}
r_H\approx2M(v)(1+4M'(v))\ .
\ee
Consider now the ingoing Vaidya metric:
\be\label{vaidya2}
ds^2=-\left(1-\dfrac{2M(v)}{r}\right)dv^2+2dv\,dr+r^2d\Omega^2
\ee
Let $\epsilon\equiv r-r_H$, where $r_H=r_H(v)$ is the location of the event horizon. Then $d\epsilon=dr-r_H'dv$ and the metric \eqref{vaidya} becomes
\be
ds^2=-\left(1-\dfrac{2M(v)}{r}-2r_H'\right)dv^2+2dvd\epsilon+r^2d\Omega^2\ .
\ee
The event horizon is the null surface given by
\be
0=\left(1-\dfrac{2M(v)}{r}-2r_H'\right)\quad\rightarrow\quad r_H'=\dfrac{r_H-2M(v)}{2r_H}\ ,
\ee
which implies that
\be\label{rdot2}
r_H''=\dfrac{M(v) r_H-2M(v)^2-2M'(v)r_H^2}{r_H^3}
\ee
The condition \eqref{sp11} gives
\be
r_H=\dfrac{M(v)-\sqrt{M(v)^2-16M(v)^2M'(v)}}{4M'(v)}\ ,
\ee
which to leading order in $L=M'(v)$ is
\be
r_H\approx2M(v)(1+4M'(v))\ ,
\ee
which is precisely the same as \eqref{reh2}. Therefore to leading order in $L=M'(v)$ and near the apparent horizon $r_{AH}=2M$, the behaviour of radial null rays in a general spacetime given by \eqref{met} is well approximated by the {\it linear} ingoing Vaidya metric \eqref{vaidya2}. This is in agreement with recent work which has shown that the requirement of regularity at the apparent horizon and consistency with the first law of black hole thermodynamics constrains the near-horizon geometry to be approximately linear Vaidya \cite{Dahal2023}.

\section{Vaidya is Conformal to Rindler space}\label{sec3}

We next discuss the role of conformal symmetry and Rindler space in determining the Hawking temperature of the Vaida spacetime. Fundamental to general relativity (indeed all metric theories of gravity) is the notion of the local equivalence of a uniform gravitational field and an accelerated reference frame in Minkowski spacetime. This manifestation of the Einstein equivalence principle (EEP) arises from local Lorentz invariance and the universal coupling of all non-gravitational fields to the same symmetric, rank-two tensor field, both of which enjoy a wealth of observational evidence. The precise statement of the EEP is that for any future-directed timelike geodesic curve $\gamma(t):I\rightarrow\mathcal{M}$ in a spacetime $(\mathcal{M},g)$, there exist Fermi coordinates $\{x_0,x_1,x_2,x_3\}$ at a point $p$ along $\gamma$ in a neighbourhood $\mathcal{U}\ni p$ for which $\Gamma^a_{\ bc}=0$ for all of $\gamma\in\mathcal{U}$. This establishes an important relationship between accelerated observers and gravitational phenomena, as non-extremal horizons are generally locally Rindler, and pointlike uniformily accelerated observers perceive the Minkowski vacuum state as a thermal bath at the Unruh temperature $T_U=a/2\pi $ \cite{unruh1976}. For a Schwarzschild black hole with surface gravity $\kappa=1/4M$ the near-horizon geometry is just the Rindler geometry with acceleration parameter $a=\kappa^{-1}$, and the resulting temperature $T_U$ exactly coincides with the Hawking temperature $T_H=1/8\pi M $ \cite{hawking1975}.
\\

This suggests a formal relationship between the Hawking and Unruh effects, which nevertheless have distinct origins. Indeed the physical interpretation of $T_U$ as the temperature of a thermal bath cannot easily be extended to macroscopic objects coupled to the vacuum, for which spatial homogeneity can no longer be maintained \cite{buchholz2015,lima2019}. Moreover, while extracting global properties of Hawking radiation from the near-horizon geometry in this way seems natural---the Hawking effect is ordinarily thought to be of near-horizon origin---Hawking quanta have wavelengths $\sim M$ and can in no way be localized to any region `near' the horizon. Despite many investigations into the matter, there still remains a lack of consensus about what physical interpretation should be assigned to the Unruh effect \cite{buchholz2013}.
\\

Regardless, in many cases Rindler geometry can still be concretely used to extract the late-time Hawking temperature associated to radiating black hole spacetimes through their near-horizon geometries. This is because the spectrum of late-time Hawking radiation depends only on the black hole having an approximately constant surface gravity $\kappa_c$ and a would-be apparent horizon over some appropriate timescale \cite{visser2003,barcelo2011}. Through suitable coordinate transformations this $\kappa_c$ is directly related to the acceleration along orbits of the boost operator in the Rindler decomposition of Minkowski space.  As we will show, the Vaidya metric is conformally related to the Rindler metric, providing a prescription for determining the Hawking temperature of dynamical black holes which are modelled by \eqref{vaidya2}. To demonstrate this, we begin by writing the Vaidya metric as
\be\begin{aligned}\label{vaidya}
ds^2&= - f(v,r) dv^2+ 2 dv dr+ r^2 d\Omega^2\ ,\\
 f(v,r)&\equiv 1- \frac{2 M(v)}{r}\ .
\end{aligned}\ee
Consider then a coordinate transformation of the form
\be
r= 2 M(v)(1+ \epsilon^2+ 4 M'(v))\ ,
\ee
where  $\epsilon^2 \ll 1$ and $M'(v) \ll 1$ independently. This is analogous to the near-horizon expansion of the Schwarzschild metric which leads to the Rindler metric, with an additional `dynamical' term.  To leading order in $M'(v)$ and $\epsilon^2$ we have that
\be\begin{aligned}
dr &\approx 2 M'(v) (1+ \epsilon^2+ 4 M'(v)) dv+ 4 M(v) \epsilon d\epsilon\ ,\\
&f(v,r)\approx \epsilon^2+ 4 M'(v)\ ,
\end{aligned}\ee
and to the same order of approximation the metric becomes
\be
ds^2 \approx - \epsilon^2 dv^2 + 8 M(v) \epsilon dv d\epsilon +4 M(v)^2 d\Omega^2\ , \label{rind19}
\ee
where we further assume that $\epsilon M'(v)\ll1$ and $M''(v)\ll1$. When $M(v)=\mathrm{const.}$ this is the Rindler metric in $(v,r)$--coordinates. From \eqref{rind19} it is evident that the Rindler horizon associated with the Vaidya metric is different from that of the Schwarzschild metric. For the Vaidya metric, the horizon is located at
\be
\epsilon^2 =0\quad \implies \quad r_H= 2 M(v) (1+ 4 M'(v))\ ,
\ee
while for the Schwarzschild metric it is simply at $r_H=2M$. The acceleration of the Rindler observer grows without bound as $\epsilon \to 0$, though in any case the reference frame fails to be rigid prior to reaching this limit (see e.g. \cite{buchholz2015}). 
\\

\begin{figure}
\includegraphics[width=5cm]{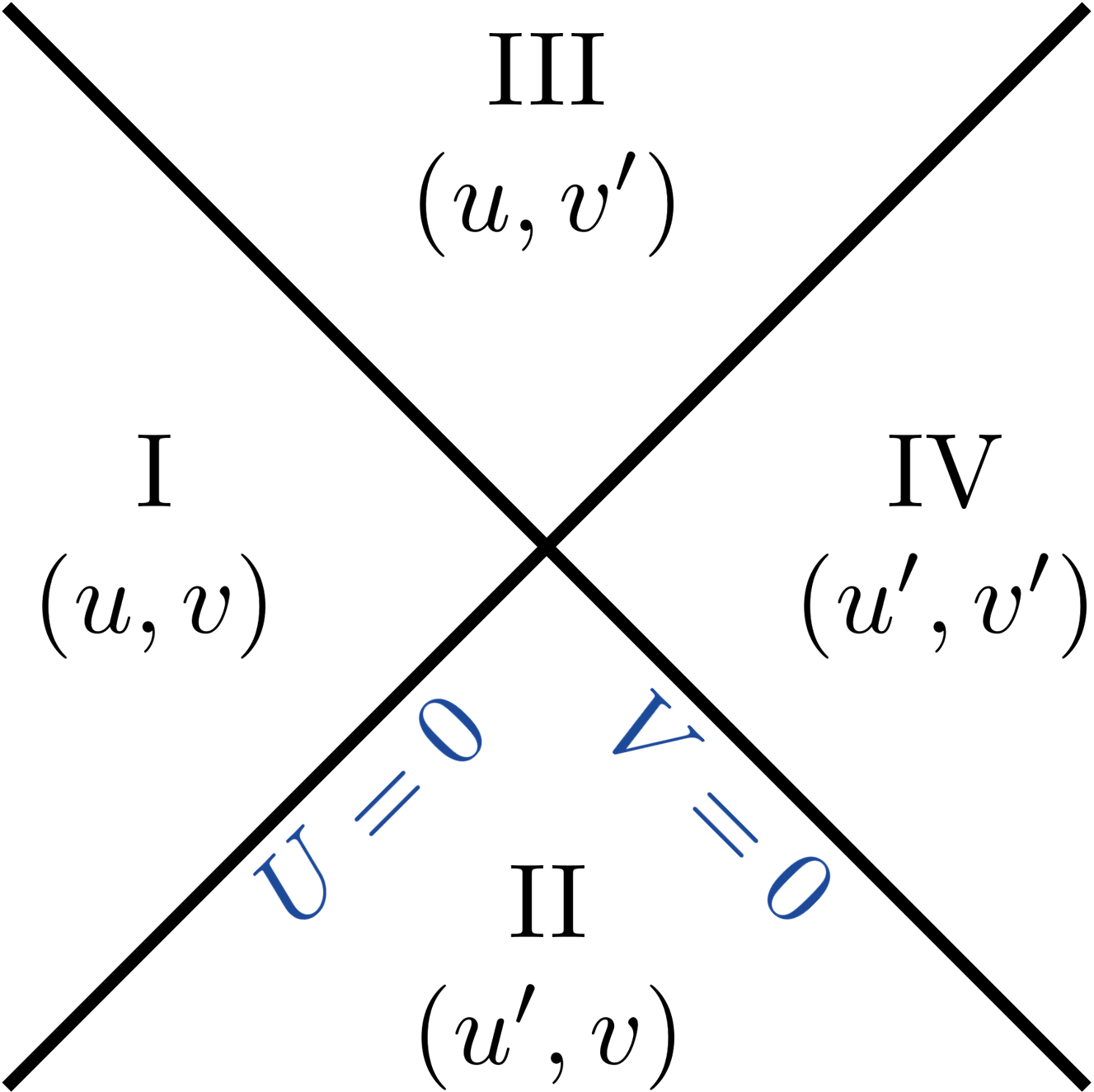}\label{fig1}
{\small \caption{The left wedge I in the Minkowski diagram represents the Rindler spacetime, which describes the frame of a uniformly accelerated observer. In this frame, the Rindler horizon separates the domain of causal contact of the accelerating observer and serves as a separatrix for orbits of the boost operator. Analytic continuation of the Rindler wedge to all four quadrants gives the complete Minkowski spacetime. The structure of this diagram is unchanged by conformal transformations.}}
\end{figure}

Now let
\be\label{rin9}
du= d\tilde{v}- \frac{2}{\epsilon} d\epsilon \ ,\quad d\tilde v\equiv \frac{dv}{ 4 M(v)}
\ee
such that \eqref{rind19} becomes
\be
ds^2= - 16 M(\tilde v)^2 \epsilon^2 du d\tilde v+ 4 M^2 d\Omega^2\ . \label{rin20}
\ee
From \eqref{rin9} this metric can be written as
\be
ds^2= -a^2 e^{\tilde v-u} du d\tilde v+ 4 M^2 d\Omega^2\ , \label{rin13}
\ee
where
\be
a= \sqrt{16 M^2 \epsilon^2 e^{u-\tilde v}}= 4 M=\kappa^{-1}\ .
\ee
For a given $v$, the acceleration parameter $a$ therefore has magnitude equal to the inverse surface gravity of the Schwarzschild spacetime. Since $a= a(v)$ we can further make the transformation
\be
dV'= a(\tilde v)^2 e^{\tilde v} d\tilde v \implies V= \frac{V'}{g(V')}= e^{v}, \quad U= e^{-u}\ , \label{ct29}
\ee
where
\be
\frac{d g(v)}{dv}+ g(v)= 16 M^2 \implies g(v)= e^{-v} \int 16 M^2 e^{v} \, dv\ .
\ee
For the stationary case, $g(v)= 16 M^2$. This transformation enables us to write the above metric as the flat spacetime metric
\be
ds^2= dU dV'+ 4 M^2 d\Omega^2= 16 M^2 dU dV+ 4 M^2 d\Omega^2\ . \label{rin16}
\ee
By expressing the metric in $(U,V)$ coordinates we can achieve the desired analytic extension, as depicted in Fig.~\ref{fig1}. This is needed because the Rindler coordinates $(u, v)$ cover only one wedge of the complete Minkowski spacetime. To cover all four quadrants, these coordinates must be extended beyond this wedge by introducing additional coordinates $(u', v')$. The corresponding $(U,V)$ for these new coordinates would be
\be
V= \frac{V'}{g(V')}= e^{-v'} \quad \textrm{for } V<0, \qquad U= e^{u'} \quad \textrm{for } U<0\ .\nonumber
\ee
To relate the metric above to the Rindler metric, let
\be
d\tilde{v}=dt+\dfrac{1}{\epsilon}d\epsilon\ .
\ee
The metric \eqref{rin20} then factorizes into an $\mathcal{S}^2$ and a two-dimensional $\mathcal{S}^1\times\text{I}\!\text{R}$ conformal Rindler metric:
\be\label{rind21}
ds^2_2= 16 M(\tilde{v})^2 (- \epsilon^2 dt^2+ d\epsilon^2)
\ee
Therefore, to leading order in the evaporation rate the near-horizon geometry of the Vaidya metric \eqref{vaidya} is accurately described by the {\it conformal} Rindler metric \eqref{rind21}. As a result the Hawking temperature associated with a {\it dynamical} spacetime represented by \eqref{vaidya} can be readily computed.

\section{The Temperature of Vaidya}\label{sec4}

We will now determine the Hawking temperature associated with the metric \eqref{rind21}, and argue that to leading order in $M'(v)$ the dynamics of the Vaidya spacetime---which well describes the near-horizon geometry of physical black holes---is reasonably approximated by a sequence of Schwarzschild solutions labelled by the time parameter $v$. Despite being often assumed, it is not obvious that this should be possible. While a naive replacement $M\rightarrow M(v)$ in the Hawking temperature
\be\label{hawkingT}
T_H=\dfrac{1}{8\pi M} \quad \rightarrow \quad \dfrac{1}{8\pi M(v)}
\ee
at first appears to be a logical generalization to the dynamical case, there are good reasons to question the validity such a replacement. For one, it is not known whether the presence of trans-Planckian modes significantly alters the near-horizon structure, which is the region responsible for the exponential redshifting of ingoing modes during collapse that leads to the Hawking spectrum in the asymptotic region. Such modes are a consequence of a dynamical metric and are absent in treatments of Hawking radiation based on the eternal Schwarzschild geometry. For two, there is a disagreement among conclusions from 1+1-dimensional models of back-reaction---some predict the formation of a macroscopic stable remnant, where a singularity on the apparent horizon develops and ceases the Hawking process, while others maintain complete evaporation \cite{susskind1992}. Effective field theory approaches to the issue, such as those studied by CGHS \cite{callan1992}, further predict that one-loop corrections responsible for quantum back-reaction arise well before the Planck scale (before the semiclassical description would break down). Hence there remains uncertainty about the role of back-reaction even in the classical description of black hole evaporation. \\

Many of the usual methods for computing the Hawking temperature are limited in their applicability, requiring static backgrounds (as in derivations based on the Euclidean path integral), thermodynamic arguments (e.g.  defining the temperature as being the quantity conjugate to the entropy variation in the first law $dE=TdS$), or a particular notion of surface gravity (whose constancy over the event horizon is assured by the strong rigidity theorem). That these methods all agree and predict the value \eqref{hawkingT} for the static Schwarzschild geometry is an encouraging demonstration of the robustness of the Hawking process, but says little about its extension to the dynamical case. Once the geometry becomes non-stationary, different notions of surface gravity become non-degenerate, defining a dynamical entropy becomes contentious, and the back-reaction problem (at some stage) is realized in full form.
\\

Some approaches to addressing the problem of Hawking radiation in time-dependent backgrounds suggest that the dynamics of the Vaidya metric cannot be decoupled from the static Schwarzschild behaviour, leading to dynamical corrections to the Hawking temperature \cite{jing-yi2006}. As we will show, this will not be the case for slowly evaporating black holes described by the Vaidya metric, even when the effects of back-reaction are implicitly included in $T_{\mu\nu}$. To do so, we appeal to relationships between the various conformally related spacetimes which enter the description of an evaporating black hole.

\subsection{Conformal Vacua}

We have shown that the near-horizon geometry of a generic evaporating black hole spacetime is given by the Vaidya metric, and that this metric is conformally related to the Rindler metric. We now argue that this can be used to determine the temperature of Hawking radiation in the Vaidya background. 
\\

A massless scalar field $\phi$ propagating in a conformally flat spacetime $g_{\mu\nu}$ is trivially conformally invariant in $D=2$, while in higher dimensions conformal invariance of $\phi$ requires that the conformal coupling $\xi$ satisfy
\be
\xi=\frac{(D-2)}{4(D-1)}\ ,
\ee
where the conformally flat spacetime is related to the Minkowski spacetime through
\be\label{conformal1}
g_{\mu\nu}=\Omega^2\eta_{\mu\nu}\ .
\ee
Under a transformation \eqref{conformal1} the scalar field $\phi$ transforms as
\be
\phi'=\Omega^{(D-2)/2}\phi\ , 
\ee
such that for $D=2$, fields in the conformally related spacetimes are equivalent (higher spin fields transform analogously with a spin-dependent conformal weight). As a result, positive frequency mode solutions $u_k$ for the field $\phi$ are the same as for $\phi'$, as are the creation and annihilation operators $(a_k^\dagger,a_k)$ they define. The unique vacuum state defined by $a_k\ket{0}=0$ is the same for both $\phi$ and $\phi'$ and is called the {\it conformal vacuum}. Two-point correlation functions evaluated with respect to $\ket{0}$ in the Minkowski ($\eta$) and conformal  Minkowski ($g$) spacetimes are likewise related (see \cite{birrell1994}):
\be
G_\eta(x, x^{\prime})=\Omega^{(D-2) / 2}(x) G_g(x, x^{\prime}) \Omega^{(D-2) / 2}(x^{\prime})
\ee
An analogous relationship exists between Rindler and conformal Rindler spacetimes. This existence of the conformal vacuum will be used below to show that conformal Rindler vacuum appears as a bath of thermal radiation to the Minkowski observer. Moreover, the temperature of this bath is same temperature (with appropriate rescaling) as that of Rindler vacuum to the Minkowski observer.\\

In the coordinates introduced above, the action for a free massless Klein-Gordon field in $D=2$ is
\begin{align}\label{action}
    {\cal S}&= \frac{1}{2}\!\! \int d^2 x \sqrt{-g} \ \partial_a\phi \partial^b\phi \nonumber\\
    &= - \int\!\! du dv\ \partial_u\phi \partial_{v}\phi= \int\!\! dU dV\ \partial_U\phi\, \partial_V\phi\ .
\end{align}
The Lagrangian corresponding to this action is invariant under a conformal transformation of the form \eqref{conformal1}, which for the Rindler metric \eqref{rind21} implies that
\begin{align}
    \tilde g_{a b}\to g_{a b}&= 16 M(v)^2 \tilde g_{a b}\ ,\\
    \tilde\phi \to \phi&= \frac{1}{4 M(v)} \tilde\phi\ .
\end{align}
The equation of motion arising from the variation of \eqref{action} with respect to $\phi$, and its conformal transformation, are thus given by
\begin{equation}
    \square\phi= (4 M(v))^{-3} \tilde\square\tilde\phi=0 \ ,
\end{equation}
implying that the fields $\tilde\phi$ and $\phi$ simultaneously solve their corresponding equation of motion. $\tilde\square\tilde\phi=0$ is the wave equation in Rindler/Minkowski spacetime, and admits two independent solutions $\tilde f_k$ and $\tilde f_k^*$ for the mode functions. Under a conformal transformation to the Vaidya spacetime, these mode functions become $f_k= \tilde f_k/(4 M(v))$ and $f_k^*= \tilde f_k^*/(4 M(v))$. That they are decomposed into positive and negative frequency unambiguously (due to the existence of the conformal vacuum) ensures that the particle concept and a notion of vacuum is well defined for this field and the unique vacuum state is called the conformal vacuum.\\

We now have the solutions of the field equations in terms of positive and negative frequency modes for both the conformal Rindler and conformal Minkowski spacetimes above, as well as the coordinate transformation \eqref{ct29} which relates them. These ingredients allows us to determine the temperature perceived by the conformal Rindler observer in the conformal Minkowski vacuum, in analogy with the standard Unruh effect (where it is the temperature perceived by a Rindler observer measuring the Minkowski vacuum). This is because the modes above are rescaled in the same way between Minkowski and conformal Minkowski spacetimes as they are from Rindler to conformal Rindler. The given coordinate transformation also coincides with the usual Rindler to Minkowski transformation. The temperature measured by the conformal Rindler observer in the conformal Minkowski vacuum is therefore $T= 1/2\pi$, where the factor $a$ that ordinarily appears is contained in the coordinate definition of $V'$ in \eqref{ct29}. Since the vacuum state for the Minkowski and conformal Minkowski spacetimes coincide, the temperature measured by the conformal Rindler observer in the conformal Minkowski vacuum would simply be
	\begin{equation}\label{temp}
	T= \frac{1}{2\pi} (\textrm{redshift factor})= 
	\frac{1}{8\pi M(v)}\ .
	\end{equation}
	The redshift factor appears due to the normalization of the conformal factor in the asymptotic region, which should be rescaled to approach unity there \cite{nielsen2013a}. Our result is consistent with established conformal relations for invariant free fields in stationary spacetimes, where it is known that the Hawking temperature is invariant under conformal transformations of the background. This conclusion can be reached by explicit computation of the relevant Bogoliubov coefficients, or more simply by defining a conformally invariant surface gravity $\kappa_c$ associated with the conformal Killing horizon, as done in \cite{jacobson1993d}. In the former case, Bogoliubov coefficients are computed by tracing null rays backward from the asymptotic region. Some of the associated modes scatter off of the effective potential, while the ones important for the observed Hawking flux have support largely in the near-horizon region. The preceding logic should therefore be applicable to a wide variety of dynamical spacetimes, so long as the near-horizon region exhibits a conformal invariance.

\subsection{Euclidean Methods}\label{sec4b}

That the temperature in the conformal Rindler metric should be given by \eqref{temp} can also be argued on the basis of the path integral. It is well-known that states at finite temperature in Lorentzian backgrounds have descriptions in terms of periodic states in the corresponding Euclidean sections of those backgrounds.  Formally, this arises from the fact that the Boltzmann factor for a system at finite $T=\beta^{-1}$ and time evolution operator are related by a periodic identification in imaginary time. The partition function in the canonical ensemble for a Hamiltonian $\hat{H}$ at finite temperature is
\be
\mathcal{Z}=\text{Tr }e^{-\beta \hat{H}}=\int dq \bra{q}e^{-\beta \hat{H}}\ket{q}
\ee
where the time evolution operator appears explicitly under the trace when $\beta=-i t$. In the case of gravity, the semi-classical approximation allows one to compute the leading contribution to $\mathcal{Z}$ through a Euclidean path integral over the Euclidean section of the metric in question \cite{hawking1978}, provided the Euclidean time $\tau=-it$ is periodically identified, $\tau=\tau+\beta$. One has that
\be
\mathcal{Z}=\int\!\! \mathcal{D}[g]\ e^{-I_E[g] / \hbar}\approx e^{-I_E\left[g_{c l}\right] / \hbar}\ ,
\ee
where $I_E[g]$ is the Euclidean action for the metric $g$, and $I_E\left[g_{c l}\right]$ is the saddle-point contribution from the classical metric $g_{cl}$. In this case $g_{cl}$ will be the Euclidean Schwarzschild solution, whose near-horizon geometry is given by the metric
\be\label{schw}
ds^2\approx \epsilon^2 d\tau^2+ 
 16 M^2 d\epsilon^2+4 M^2d\Omega_2^2\ ,
\ee
where $r \equiv 2M(1+\epsilon^2)$. For the path integral (and thus canonical ensemble) to be well-defined, the metrics that contribute to $\mathcal{Z}$ must be regular. The Euclidean Schwarzschild solution $\mathcal{E}$ is foliated by leaves of topology $\mathcal{S}^1\times\mathcal{S}^2$, and is nominally singular at $\epsilon=0$. In order for the proper length of the $\mathcal{S}^1$ to shrink smoothly to zero on the horizon the periodicity  $\tau$ must be such that
\be\label{period}
\tau=\tau+8\pi M \quad\implies\quad \beta=8\pi M\ . 
\ee
This eliminates the conical singularity that would generically appear in the Euclidean section, so that the metric is positive-definite. The leading contribution to the path integral from $\mathcal{E}$ is then well-defined, and the required periodicity \eqref{period} encodes the KMS condition \cite{kubo1957,martin1959a} for Green's functions on $\mathcal{E}$ . These Green's functions are analytic continuations of Green's functions defined on the Lorentzian geometry $\mathcal{M}$ which define a thermal state at finite temperature
\be
\beta^{-1}=T=\dfrac{1}{8\pi M}\ .
\ee
Importantly, the conformal factor $16M(\tilde{v})^2$ appearing in \eqref{rind21} does not alter the periodicity required for smoothness at the horizon and therefore should not alter the equilibrium temperature associated with the geometry. In fact, any conformal factor which is finite at the boundary $r\rightarrow \infty$ and can be normalized there such that $\Omega\rightarrow 1$  will define a spacetime with the same temperature. In the full path integral, it is necessary to separate metrics which contribute to the measure $\mathcal{D}[g]$ into equivalence classes under conformal transformations with the appropriate boundary conditions. If the spectrum of the wave operator $(\square-\xi R)$ is positive, the Euclidean action is well-defined and given by a pure boundary term with periodicity equal to that of the conformally related action \cite{gibbons1978}. 
\\

\section{The Kerr-Vaidya spacetime}\label{sec5}

The conformal transformation used to bring the Vaidya metric into conformally flat form can also be applied to the Kerr-Vaidya solution, which serves as a better proxy for rotating astrophysical black holes. As before, we assume slow evaporation and work to leading order in the evaporation rate. The Kerr-Vaidya metric in advanced coordinates is given by \cite{dahal2020a}
\begin{align}\label{kv}
ds^2&=-\left(1-\frac{2 M(v) r}{\rho^2}\right)dv^2+2 dv dr-2 a \sin^2\!\theta d\phi dr\nonumber\\
&\quad+ \rho^2 d\theta^2-
\frac{4 a M(v) r \sin^2\theta}{\rho^2}d\phi dv
+
\frac{{\cal A}}{\rho^2}\sin^2\theta d\phi^2\ ,\nonumber
\end{align}
where
\begin{align}
\rho^2&= r^2+ a^2 \cos^2\theta\ , \quad \Delta= r^2+ a^2- 2 M r\ ,\nonumber\\
{\cal A}&= (r^2+ a^2)^2- \Delta a^2 \sin^2\theta\ .
\end{align}
This metric can instead be written in the form
\begin{equation}
ds^2= -\frac{\Delta \rho^2\ d\tilde v^2}{(r^2+ a^2)^2} + \frac{2\rho^2\ d\tilde v dr}{r^2+ a^2} + \rho^2( d\theta^2+  \sin^2\theta d\tilde\phi^2)\ ,\nonumber
\end{equation}
where
\be\begin{aligned} \label{nc53}
d\tilde v&= \frac{(r^2+ a^2)}{\rho^2} \left(dv- a \sin^2\theta d\phi\right)\ ,\\
d\tilde\phi&= \frac{(r^2+ a^2)}{\rho^2} \left(d\phi- \frac{a}{a^2+ r^2} dv\right)\ .
\end{aligned}\ee
$d\tilde v$ and $d\tilde\phi$ are the respective analogs of $dt$ and $d\phi$ of the Schwarzschild metric. This correspondence is almost exact with one exception: $d\tilde v$ and $d\tilde\phi$ together with $d\theta$ and $dr$ form an anholonomic basis of one-forms. This means that there are no globally defined coordinates $X$ and $\tilde X$ such that $d\tilde v$ = $dX$ and $d\tilde\phi  = d\tilde X$.
Near the outer horizon $r_+= M(v)+ \sqrt{M(v)^2- a^2}$, we then make the following approximations:
\begin{align}
\Delta= \frac{\gamma^2 \tilde\epsilon^2}{4}+ \gamma h(v) &M(v)'\ ,\quad r- r_+= \frac{\gamma \tilde\epsilon^2}{4}+ h(v) M(v)'\ , \nonumber\\
\gamma&= 2 \sqrt{M(v)^2-a^2}\ ,
\end{align}
where $h(v)$ is a function determining the location of Rindler horizon. For $r= r_+ + \gamma \tilde\epsilon^2/4+ h M(v)'$, we thus have that
\begin{equation}
dr= M(v)' \left(1+ \frac{2 M(v)}{\gamma} \right) d\tilde v+ \frac{\gamma \tilde\epsilon}{2} d\tilde\epsilon\ ,
\end{equation}
and the Kerr-Vaidya metric \eqref{kv} reduces to
\begin{align}
ds^2&= -\frac{\rho_+^2}{r_+^2+a^2}\left[ \left(\frac{\gamma h}{r_+^2+a^2}- 2\left(1+\frac{2 M(v)}{\gamma}\right)\right)M(v)'\right.\nonumber\\
&\quad\left.+\frac{\gamma^2 \tilde\epsilon^2}{4(r_+^2+a^2)} \right] d\tilde v^2+ \frac{\rho_+^2 \gamma \tilde\epsilon}{r^2+ a^2} d\tilde v d\tilde\epsilon\\
&\quad+ \rho_+^2 \left(d\theta^2+  \sin^2\theta d\tilde\phi^2 \right)\ ,\nonumber
\end{align}
where only terms up to order $\tilde\epsilon^2$ and $M'(v)$ have been retained. The function $h(v)$ is now chosen to be
\begin{equation}
h= \frac{4 r_+ (r_+^2+ a^2)}{\gamma^2}\ ,
\end{equation}
which reduces the above metric to the form
\begin{equation}\label{kv2}
ds^2= -\frac{\rho_+^2 \gamma^2 \tilde\epsilon^2\ d\tilde v^2}{4(r_+^2+a^2)^2} + \frac{\rho_+^2 \gamma \tilde\epsilon\ d\tilde v d\tilde\epsilon}{r_+^2+ a^2} + \rho_+^2 \left(d\theta^2+  \sin^2\theta d\tilde\phi^2 \right),
\end{equation}
Finally, we make a further substitution
\begin{equation}
\rho d\tilde\epsilon= d\epsilon\ ,
\end{equation}
which implies that
\begin{align}
d(\rho \tilde\epsilon)&= \frac{2 r \tilde\epsilon dr- a^2 \sin2\theta \tilde\epsilon d\theta }{2\rho} + \frac{2 \rho}{\gamma \tilde\epsilon} dr \nonumber\\
&\hspace{2.5cm} - \frac{2 \rho M'(v)}{\gamma \tilde\epsilon} \left(1+ \frac{2 M(v)}{\gamma}\right) d\tilde v\nonumber \\
&\approx \frac{2 \rho}{\gamma \tilde\epsilon} dr- \frac{2 \rho M'(v)}{\gamma \tilde\epsilon} \left(1+ \frac{2 M(v)}{\gamma}\right) d\tilde v \nonumber\\
&= \rho d\tilde\epsilon= d\epsilon.
\end{align}
Substituting these relations into \eqref{kv2}, we obtain
\begin{equation}
ds^2= -\frac{\gamma^2 \epsilon^2\ d\tilde v^2}{4 (r_+^2+a^2)^2} +  \frac{\gamma\epsilon\ d\tilde v d\epsilon}{r_+^2+a^2} + \rho_+^2 (d\theta^2+ \sin^2\theta d\tilde\phi^2)\ .
\end{equation}
Just as in the spherically symmetric case, this can be written in advanced and retarded coordinates as
\begin{align}\label{kv3}
ds^2&= -\frac{\gamma\epsilon^2}{2(r_+^2+a^2)} d\tilde v \left(\frac{\gamma}{2(r_+^2+a^2)} d\tilde v- \frac{2}{\epsilon} d\epsilon\right)\\
&= -\frac{4(r_+^2+a^2)^2 \epsilon^2}{\gamma^2}\, du' dv'+ \rho_+^2 (d\theta^2+ \sin^2\theta d\tilde\phi^2)\ \nonumber,
\end{align}
where
\begin{equation}
du'= \frac{\gamma}{2(r_+^2+a^2)} d\tilde v- \frac{2}{\epsilon} d\epsilon\ , \quad dv'= \frac{\gamma^3}{8 (r_+^2+a^2)^3} d\tilde v\ .\nonumber
\end{equation}
The second line of Eq. \eqref{kv3} represents the near horizon limit of the Kerr-Vaidya metric (after series of coordinate transformations including anholonomic ones) and is seen to be conformal to the Minkowski metric. Therefore, as for the spherically symmetric Vaidya case above, the temperature corresponding to this metric for a stationary distant observer (accounting for the redshift factor) is simply
\begin{equation}
T= \frac{1}{2\pi}\frac{\gamma}{2(r_+^2+a^2)}= \frac{\sqrt{M(v)^2-a^2}}{2\pi (r_+^2+a^2)}.
\end{equation}
This is the Hawking temperature associated with the Kerr-Vaidya metric \eqref{kv}.

\section{Discussion}\label{sec6}

In this work we have shown that regularity assumptions at the apparent horizon---which are necessary for the persistence of the evaporation process---imply a near-horizon geometry for evaporating black holes that is described by the ingoing Vaidya metric. While such a metric is insufficient to capture the physical features of the entire evaporation process, it nevertheless serves as a useful model-building tool which accurately presents the ingoing (negative energy) flux near the horizon associated with the Hawking process. A complete model needs to also account for the outgoing flux observed in the far region, as done in \cite{hiscock1981}. We also show that a particular coordinate transformation allows this near-horizon geometry to be uniquely expressed in terms of the {\it linear} ingoing Vaidya metric to leading order in the evaporation rate, i.e. for slowly evaporating black holes. We demonstrate that this metric can be brought into a conformally static form, and by appealing to the equivalence of the conformal vacua in the Minkowski/Rindler and conformal Minkowski/Rindler spacetimes, we are able to show that the Hawking temperature associated with the Vaidya metric is simply
\be
T=\dfrac{1}{8\pi M(v)}\ ,
\ee
again to leading order in the evaporation rate $M'(v)$. What we have done is analogous to derivations of the Hawking temperature which rely on the near-horizon expansion of the static Schwarzschild geometry being a Rindler geometry. In our case however, the corresponding near-horizon expansion of the Vaidya metric gives rise to a conformal Rindler geometry. That the evaporation rate $M'(v)$ does not appear in $T$ shows that (small) dynamical effects do not contribute significantly to the Hawking process, and therefore the evaporation of the black hole can be appropriately modelled by a quasistatic sequence of Schwarzschild metrics with changing mass parameter $M(v)$. That this is the case is not obvious given the present uncertainty regarding the effects of back-reaction for evaporating black holes. The existence of a well-defined temperature for the Vaidya metric is likely to further play a role in generalizing the laws of black hole mechanics to dynamical spacetimes.
\\

To the best of our knowledge, the work we present here also represents the first explicit calculation of the Rindler horizon associated with a non-stationary spacetime. Our investigation of the Vaidya spacetime has revealed that the location of the Rindler horizon coincides precisely with the location of the separatrix. Furthermore, we have found that this location plays a critical role in determining the nature of black hole evaporation in this particular example. While our findings are intriguing, there remains much work to be done to fully understand the general dynamical case. We make steps towards this by generalizing the conformal transformation to the dynamical rotating case of Kerr-Vaidya black holes. Future work will focus on investigating whether such conformal transformations are limited in applicability to Vaidya and Kerr-Vaidya spacetimes, or for a wider class of examples. If so, such transformations should prove useful for studying a variety of models of evaporating astrophysical black holes.

\section*{Acknowledgements}

We would like to thank Daniel R.\ Terno for useful comments. P.K.D. is supported by an International Macquarie University Research Excellence Scholarship. F.S. is funded by the ARC Discovery Project Grant No. DP210101279.

\section*{Appendix}\label{AppA}

In this appendix, we demonstrate how regularity assumptions lead to the approximations required in the body to bring the general spherically symmetric metric \eqref{met} into the form \eqref{met1} near the apparent horizon. We begin with the general spherically symmetric metric in both Schwarzschild and ingoing null coordinates
\begin{align}
ds^2&=-e^{2\tilde{\phi}(t,r)}f(t,r)\,dt^2+\dfrac{dr^2}{f(t,r)}+r^2d\Omega^2\ ,\\
ds^2&=-e^{2\phi(v,r)}f(v,r)\,dv^2+2e^{\phi(v,r)}dv\,dr+r^2d\Omega^2\ ,\nonumber
\end{align}
where the metric function is invariant on the apparent horizon, namely $f(t,r)=f(v,r)$ at $r=r_{AH}$. Between these two coordinate systems, the stress-energy components are related through $T_{\mu\nu}={\Lambda_{\mu}}^{\alpha}{\Lambda_{\nu}}^{\beta}T_{\alpha\beta}$. We first define the rescaled stress-energy tensor components
\be
\tau_t\equiv e^{-2\phi}T_{tt}\ ,\quad {\tau^r}_t\equiv e^{-\phi}f T_{tr}\ ,\quad\tau^{rr}\equiv f^2T_{rr}\ . 
\ee
Then, we have that
\begin{align}
T_{vv}&=\frac{\partial x^{\alpha}}{\partial v}\frac{\partial x^{\beta}}{\partial v}T_{\alpha\beta}\nonumber\\
&=e^{2\tilde{\phi}}\,\tau_t\ ,
\end{align}
where the second two terms in the second line vanish since $\partial_v r=0$. Likewise
\begin{align}
T_{vr}&= \frac{\partial x^{\alpha}}{\partial v}\frac{\partial x^{\beta}}{\partial r}T_{\alpha\beta}\nonumber\\
&=-\dfrac{e^{\tilde{\phi}}}{f}\tau_t+\dfrac{e^{\tilde{\phi}}}{f}{\tau^r}_t\ ,
\end{align}
and
\begin{align}
T_{rr}&= \frac{\partial x^{\alpha}}{\partial r}\frac{\partial x^{\beta}}{\partial r}T_{\alpha\beta}\nonumber\\
&=\dfrac{\tau_t-2{\tau^r}_t+\tau^{rr}}{f^2}\ . \label{rcd70}
\end{align}
As discussed in Refs. \cite{mann2022,Dahal2023}, the requirement that the curvature invariants such as $T_{\mu\nu}T^{\mu\nu}$ are finite on the apparent horizon gives two distinct forms of the stress-energy tensor. First, we will consider the one for which 
\be\begin{aligned}\label{emt72}
\tau_t&=-\Upsilon^2+ e_{12} f(t,r)\ , \quad  \\
\tau_t^{~r}&=-\Upsilon^2+ \phi_{12} f(t,r)\ ,\\
\tau^r&=-\Upsilon^2 + p_{12} f(t,r)\ ,
\end{aligned}\ee
up to sub-leading order. Here $\Upsilon^2$ is some real time-dependent function representing the leading contribution to the stress-energy tensor components in the near-horizon expansion, and $e_{12}$, $p_{12}$ and $\phi_{12}$ are sub-leading contributions satisfying $\phi_{12}= (e_{12}+p_{12})/2$. This is referred to as the $k=0$ solution in above references (also note that in these references it is shown that $\tau_t^{~r}<0$ implies an evaporating black hole solution).
From the second Einstein equation \eqref{efe2} we thus have
\be
\dfrac{\partial \phi}{\partial r}\propto g_1(v)\ , \label{rcd72}
\ee
which can be integrated to give
\be
\phi= h_1(v)+ g_1(v) f+ ...
\ee
Therefore, we can always rescale the time variable such that $h_1(v)=0$, and thus $\phi=0$ to leading order. The fact that $\partial_r \phi\propto g_1(v)$ and $\partial_v \phi\propto M'(v)$ ensures that we can neglect the terms $(1-2M/r)^2 \partial_r \phi$ and $(1-2M/r) \partial_v \phi$ in Eq. \eqref{sp11}. 
\\

A similar computation establishes that $\partial_r M \approx0$. We begin with
\begin{align}
{T^v}_v&=g^{vv}T_{vv}+g^{vr}T_{rv} \nonumber\\
&=\dfrac{{\tau^r}_t-\tau_t}{f}= \phi_{12}-e_{12}\ .
\end{align}
Then from \eqref{efe1} we have that
\be
\dfrac{\partial M}{\partial r} \propto g_2(v) \implies M= h_2(v)+ g_2(v) f+ ...
\ee
as desired. If the apparent horizon is taken to be the relevant surface for which the first law is formulated, then the area being $A=4\pi r_{AH}^2$ and the MSH mass being $M=r_{AH}/2$ requires that $g_2(v)=0$ for the identification of $\kappa$ with the Hayward--Kodama surface gravity~\cite{Dahal2023}. $\partial_r M(v,r)\approx0$ to leading order in the evaporation rate and near $r=r_{AH}$ implies the spatial uniformity of $M$ near the apparent horizon. If the first law is not invoked, one must take as an assumption the spatial uniformity of $M$.
\\

We can also consider the so-called $k=1$ solutions with
\begin{equation}
    \tau_t= E f(t,r)\ , \quad    \tau_t^{~r}= \Phi f(t,r)\ , \quad \tau^r= P f(t,r)\ ,
\end{equation}
with $E=-P$ and $\Phi=0$ for non-extreme (non-zero volume of trapped region) dynamical black holes. In the leading order approximation, the $k=1$ solution can be viewed as the static limit of the $k=0$ solution. This can be seen from the fact that substituting $\Upsilon^2=0$ into the stress-energy tensor of $k=0$ solution of Eq.~\eqref{emt72} gives the stress-energy tensor of the $k=1$ solution (after a redefinition of coefficients).

\newpage{\pagestyle{empty}\cleardoublepage}

\end{document}